\documentclass{article}

\usepackage{graphicx}

\def\lsi{\raise0.3ex\hbox{$<$\kern-0.75em\raise-1.1ex\hbox{$\sim$}}}
\def\gsi{\raise0.3ex\hbox{$>$\kern-0.75em\raise-1.1ex\hbox{$\sim$}}}
\newcommand{\lsim}{\mathop{\lsi}}
\newcommand{\gsim}{\mathop{\gsi}}
\newcommand{\aob}{\mbox{\footnotesize{a,obs}}}
\newcommand{\asc}{\mbox{\footnotesize{a,src}}}
\newcommand{\sob}{\mbox{\footnotesize{s,obs}}}
\newcommand{\ssc}{\mbox{\footnotesize{s,src}}}
\newcommand{\src}{\mbox{\footnotesize{src}}}
\newcommand{\obs}{\mbox{\footnotesize{obs}}}
\newcommand{\fa}{\mbox{\footnotesize{a}}}
\newcommand{\fs}{\mbox{\footnotesize{s}}}
\newcommand{\fm}{\mbox{\footnotesize{max}}}

\begin{document}
\title{Solar constraints on new couplings between electromagnetism and gravity}
\author{S. K. Solanki, O. Preuss \thanks{email: solanki@linmpi.mpg.de, opreuss@physik.uni-bielefeld.de}\\
        Max-Planck-Institut f\"ur Aeronomie\\
        D-37191 Katlenburg-Lindau, Germany\\[0.5cm]
        M. P. Haugan\\ 
        Purdue University 1396\\ West Lafayette Indiana 47907\\[0.5cm]	
	A. Gandorfer, H. P. Povel, P. Steiner, 
        K. Stucki \\
	Department of Physics \\
        Institute of Astronomy \\
        ETH-Zentrum \\ CH-8092 Z\"urich, Switzerland\\[0.5cm]
	P. N. Bernasconi\\
        Space Department \\
        Johns Hopkins University \\ Applied Physics Laboratory\\
        11100 Johns Hopkins RD\\ Laurel MD 20723-6099\\[0.5cm]
	D. Soltau\\
        Kiepenheuer-Institut f\"ur Sonnenphysik\\ D-79104 Freiburg, Germany}
\date{\today}

\maketitle
\newpage

\begin{abstract}
The unification of quantum field theory and general relativity is a fundamental
goal of modern physics. In many cases, theoretical efforts to achieve this 
goal introduce auxiliary gravitational fields, ones in addition to the 
familiar symmetric second-rank tensor potential of general relativity, and 
lead to nonmetric theories because of direct couplings between these 
auxiliary fields and matter. Here, we consider an example of a metric-affine 
gauge theory of gravity in which torsion couples nonminimally to the 
electromagnetic field. This coupling causes a phase difference to accumulate 
between different polarization states of light as they propagate through the 
metric-affine gravitational field. Solar spectropolarimetric observations are 
reported and used to set strong constraints on the relevant coupling constant 
$k$: $k^2 < (2.5\,\mbox{km})^2$.
\end{abstract}

\section{Introduction}
  The quest for a complete and self-consistent unification of quantum field 
  theory with the theory of general relativity remains one of the most 
  important unsolved problems in modern theoretical physics.  Its solution 
  promises an extension of the standard model of particle physics that 
  encompasses all fundamental interactions and accounts for everything 
  from the evolution of the early universe to black hole physics.  This 
  possibility continues to inspire work on such theories even after decades 
  of effort.  Clearly, one can expect candidate theories to predict {\it new 
  physics}, that is, phenomena beyond the scope of general relativity and 
  the standard model.  The birefringence we study here is an example.  

  Quantum gravity research is unusual in that it is driven by the need to 
  overcome extraordinary conceptual and mathematical difficulties in 
  formulating a consistent theory rather than by an accumulation of 
  experimental evidence that could inform ones choices among theoretical 
  alternatives.  Exploring such alternatives during the past several decades 
  has produced many theories of gravity and, thus, a need for an overarching 
  theoretical framework within which these alternatives can be compared and 
  classified.  The Dicke framework, defined in Appendix 4 of Dicke's 1964 
  Les Houches lectures \cite{dicke}, can be seen as the basis for a number of 
  formalisms encompassing Lagrangian-based local field theories of gravity.  
  It demands that gravity be associated with one or more fields of 
  tensorial character.  Consequently, alternatives to general relativity 
  generally feature one or more gravitational fields in addition to the 
  usual second-rank 
  symmetric tensor potential. The distinction between metric and nonmetric 
  alternatives often turns on the manner in which matter couples to such 
  additional fields. The purely geometrical character of general relativity 
  and other metric theories is a consequence of the familiar minimal coupling 
  of matter fields to a single symmetric second-rank tensor gravitational 
  field.  Nonmetric theories deviate from this pattern, typically including 
  direct couplings between matter and auxiliary gravitational fields.  The 
  Dicke framework encompasses nonmetric as well as metric theories of gravity 
  and, therefore, general relativity as a particular example for a metric 
  theory. In this paper we focus on observable consequences of a nonminimal 
  coupling between a gravitational torsion field and the electromagnetic 
  field.    
  
  An appropriate framework for the analysis of electrodynamics in a background
  gravitational field is given by the $\chi g$-formalism, invented by W.T. Ni
  in 1977 \cite{ni77}. The $\chi$ of its name refers to a tensor density which 
  provides a phenomenological representation of gravitational fields. The
  structure of the $\chi g$-formalism is in agreement with the basic 
  assumptions and constraints of the Dicke framework. Demanding 
  electromagnetic gauge invariance and linearity of the electromagnetic field 
  equations, the most general Lagrangian density governing source-free 
  electromagnetic field dynamics is 
  \begin{equation}
    {\cal L}_{\mbox{\footnotesize{NG}}} = -\frac{1}{16\pi}\chi^{\alpha\beta\gamma\delta}
    	                                  F_{\alpha\beta}F_{\gamma\delta} \quad .      
  \end{equation}
  The independent components of the tensor density 
  $\chi^{\alpha\beta\gamma\delta}$ comprise 21 phenomenological gravitational 
  potentials that allow one to represent gravitational fields in a very 
  broad class of nonmetric theories. In 1984 Ni \cite{ni84} noted that 
  theories encompassed by this formalism can predict birefringence and used 
  pulsar polarization observations to constrain this possibility.  Perhaps 
  because no specific theory predicting birefringence was recognized at that 
  time,  the significance of Ni's result was initially overlooked.   
  
  This situation changed when J.W. Moffat published a revised version of 
  his non-symmetric gravitation theory (NGT) \cite{mof79} in 1990. One year 
  later Gabriel {\it et al}. showed that NGT predicts a polarization-dependent 
  speed of light as a consequence of a violation of the Einstein equivalence 
  principle and imposed the first sharp constraints on the magnitude of this 
  birefringence \cite{gab91,gab91a,gab91b} (in this context see also 
  Solanki \& Haugan 1996 \cite{sh96} and Solanki et al. 1999 \cite{s99}). In 
  contrast to these earlier investigations, the present paper is not 
  concerned with NGT. It focuses on another alternative to general relativity 
  drawn from the class of metric-affine theories of gravity (MAG) \cite{h95}, 
  in part, because an array of technical difficulties have challenged the 
  viability of NGT \cite{d92,c98}. We show that birefringence is a generic 
  feature of MAG in the case of nonminimal coupling between electromagnetism 
  and gravity, specifically torsion, and use solar data to set sharp constraints 
  on a relevant coupling constant. We hope that this example will prompt further 
  work on gravity-induced birefringence predicted by metric-affine theories 
  and other alternatives to general relativity. In that regard, see \cite{r03,ih03}.  
  
  It is important to note that the significance of such constraints goes
  far beyond testing versions of NGT or MAG. These are merely concrete 
  examples of theories that predict gravity-induced birefringence, a 
  phenomenon shown by Haugan and Kauffmann \cite{hk95} to be predicted by 
  a broad class of the nonmetric theories encompassed by the 
  $\chi g$-formalism.  Here we make use of the formalism they invented to 
  compute the effect of gravity-induced birefringence using the $\chi g$ 
  representation of any gravitational field. We emphasize that observations 
  constraining the strength of such birefringence complement more familiar 
  tests of the Einstein equivalence principle like the E\"otv\"os, 
  gravitational redshift and Hughes-Drever experiments, which currently
  do not bound our special nonminimal coupling. The reason is that we focus 
  on a quite novel coupling between electromagnetic fields and torsion which
  was not taken into account in the analysis of these classical experiments. 
  However, since our new coupling also violates local Lorentz invariance 
  as one can see from Eq.~(\ref{lem}), it is indeed thinkable that E\"otv\"os, 
  Hughes-Drever or redshift experiments could also constrain this new possibility.

  We also report on new solar polarization data and the constraints on such 
  gravity-induced birefringence obtained therewith. In section III.A we 
  describe the new profile difference technique we use to search for 
  evidence of birefringence. In section III.B we review the Stokes 
  asymmetry technique proposed previously by Solanki and Haugan (1996) 
  \cite{sh96}. In section IV we describe the solar data we analyze and 
  the observations that produced them. The analysis of these data and the 
  constraints we infer on gravity-induced birefringence are discussed in 
  section V. We quote constraints on the Sun's NGT charge, $l_\odot ^2$, 
  purely as a figure of merit that can be compared to prior constraints,
  but focus on constraints on a metric-affine parameter, $k^2$, defined 
  in the next section.  
  
\section{Gravity-induced birefringence in metric-affine theories}
  An empirically adequate account of gravitation is given today 
  by general relativity which predicts vanishing torsion and 
  vanishing nonmetricity (covariant derivative of the metric).   
  This account is, so far, in complete agreement with observational 
  results. However, general relativity is a classical theory. The 
  desire for a quantum mechanical account of gravitation requires 
  a more general framework.
  
  One possibility is to formulate gravity as a gauge theory of an 
  underlying local spacetime group. Metric-affine gravity (MAG) 
  is based on the assumption that affine transformations are the 
  gauge group.  It is the most general canonical gauge theory of 
  gravity \cite{h95}.
  
  Metric-affine theories use a second-rank symmetric tensor field, a 
  co-frame field and a connection one-form field to represent 
  gravitational potentials.  Although the symmetric tensor is referred 
  to as the metric, metric-affine theories are generally nonmetric.  Other 
  gravitational potentials, specifically the torsion and nonmetricity fields 
  extracted from the connection, couple directly to matter.   Nonmetric 
  couplings to the electromagnetic field can lead to gravity-induced 
  birefringence. 
  
  In this section we will carefully examine the possible coupling
  of the electromagnetic field to gravity within the framework of MAG.
  Under the assumption of electric charge and magnetic flux conservation
  Puntigam et al. \cite{p97} showed that the conventional formulation of 
  Maxwell-Lorentz electrodynamics in a metric-affine gravitational field 
  predicts that the propagation of light is not influenced by the presence 
  of post-Riemannian geometric fields like torsion $T^{\alpha}$ or 
  nonmetricity $Q_{\alpha\beta}$. However, ``admissible'' nonminimal 
  coupling possibilities that lead to birefringence do exist since one can 
  modify the Maxwell-Lorentz spacetime relation $H=\lambda_0\,^{\star}F$. 
  Couplings of this kind respect gauge invariance and, 
  as a consequence, electric charge as well as magnetic flux conservation.
  In order to make quantitative predictions about electromagnetic field 
  dynamics in a metric-affine background field using this nonminimal coupling 
  scheme, a specific additional Lagrangian density is needed. We consider 
  \begin{equation}\label{lem}
    L_{EM} = k^2 *(T_\alpha \wedge F)\,T^\alpha \wedge F, 
  \end{equation}
  consistent with the suppositions made above.
  Here, $k$ is a coupling constant with units of length, $*$ denotes the 
  Hodge dual, $T$ denotes the torsion and $F$ the electromagnetic field. 
  Our intention is to set strong limits on $k^2$ and, so, to decide about the
  physical relevance of the gravity-induced birefringence generated by the 
  coupling (\ref{lem}). Currently there are no expectations from the theory
  side of what a reasonable value for $k^2$ might be. The additional Lagrangian
  (\ref{lem}) can be written as
  \begin{equation}
    \delta \! {\cal L}_{EM} = 
    \delta \! \chi^{\alpha \beta \gamma \delta} F_{\alpha \beta} F_{\gamma \delta}
    \quad , 
  \end{equation}
  so that we can make use of the general formalism developed by 
  Haugan \& Kauffmann in \cite{hk95} to infer the consequences of  
  the nonminimal coupling (\ref{lem}) on the propagation of light through a 
  metric-affine field. This means that our objective is to compute the 
  fractional difference in the propagation speed of linear polarization states 
  \begin{equation}\label{deltac}
    \frac{\delta c}{c} = \frac{1}{2}\sqrt{({\cal A} - {\cal C})^2 + 4{\cal B}^2} \quad .
  \end{equation}  
  that are singled out by a solar torsion field. The coefficients ${\cal A}$, 
  ${\cal B}$ and ${\cal C}$ depend on the location in space-time and on the direction 
  in which the wave propagates.    
  
  Since we are going to search for birefringence in the essentially static, 
  spherically symmetric gravitational field of the Sun, we are interested 
  in static and spherically symmetric solutions of the metric-affine field 
  equations for torsion. One such solution was given by Tresguerres in 
  1995 \cite{t95,t95a}.  It can be split into nonmetricity dependent and 
  independent parts. The latter, which is assumed to couple to the
  electromagnetic field via (\ref{lem}), is given by
  \begin{equation}\label{t_sol}
    T^{\alpha}=k_0\left[\frac{1}{r}(\theta^0-\theta^1)+
    \frac{m}{r^2}(\theta^0+\theta^1)\right]\wedge \theta^{\alpha}\quad ,
  \end{equation}
  with torsion mass $m$ and with $k_0=1$ for $\alpha=0,1$ or $k_0=-1/2$ for 
  $\alpha=2,3$, respectively. In this equation we have dropped the dilatation charge
  $N_0$ from Tresguerres' original solution because it vanishes if the nonmetricity
  field does. Also the small observed value of the cosmological constant $\Lambda$
  means that effects of its term can be neglected on galactic and smaller scales.
  
  Equation (\ref{t_sol}) is consistent with the most general 
  static, spherically symmetric form for a torsion field \cite{t95} 
  \begin{eqnarray}
    T^0 &=& \alpha(r)\,\theta^0\wedge\theta^1
            +\tilde{\alpha}(r)\,\theta^2\wedge\theta^3\quad , \\
    T^1 &=& \beta(r)\,\theta^0\wedge\theta^1 
            +\tilde{\beta}(r)\,\theta^2\wedge\theta^3\quad , \\
    T^2 &=& \gamma_{(1)}\,\theta^0\wedge\theta^2
            +\gamma_{(2)}\,\theta^0\wedge\theta^3 \nonumber\\
            &&{}+\gamma_{(3)}\,\theta^1\wedge\theta^2 
            +\gamma_{(4)}\,\theta^1\wedge\theta^3 \quad , \\
    T^3 &=& \gamma_{(1)}\,\theta^0\wedge\theta^3
            -\gamma_{(2)}\,\theta^0\wedge\theta^2 \nonumber\\
            &&{}+\gamma_{(3)}\,\theta^1\wedge\theta^3 
            -\gamma_{(4)}\,\theta^1\wedge\theta^2 \quad .       	       	    
  \end{eqnarray}
  The solution (\ref{t_sol}) is a special case having 
  $\tilde{\alpha}(r)=\tilde{\beta}(r)=\gamma_{(2)}=\gamma_{(4)}=0$. 

  Plugging this general representation of a spherically symmetric torsion 
  field into the Lagrangian density (\ref{lem}) yields
  \begin{eqnarray}
    \delta \! {\cal L}_{EM} &=& 
    k^2 \lbrace (\alpha^2 - \beta^2) B_1^2 
    -(\gamma_{(1)}^2 + \gamma_{(4)}^2) [B_2^2 + B_3^2]\nonumber \\ 
    &-&(\gamma_{(3)}^2 + \gamma_{(4)}^2) [E_2^2 + E_3^2] \\
    &+& 2(\gamma_{(1)} \gamma_{(4)} - \gamma_{(2)} \gamma_{(3)})[B_2 E_2 + B_3 E_3]\nonumber \\ 
    &+& 2(\gamma_{(1)} \gamma_{(3)} + \gamma_{(2)} \gamma_{(4)})[B_3 E_2 - B_2 E_3]\rbrace \nonumber
    \quad ,
  \end{eqnarray}
  where $E$ and $B$ refer in the usual way to the electric and magnetic 
  components of $F$ in Tresguerres' coordinates. Note that in his 
  notation the three-vector index 1 refers to the radial direction. 
  In terms of the components of the $\xi$, $\zeta$ and $\gamma$ tensors 
  of reference \cite{hk95}, the corresponding Lagrangian expression is 
  \begin{eqnarray}
    \delta \! {\cal L}_{EM} &=& 
    \zeta_{11} B_1^2 + \zeta_{22} B_2^2 + \zeta_{33} B_3^2 
    - \xi_{22} E_2^2 -\xi_{33} E_3^2 \\
    &+& 2 \lbrace \gamma_{22} E_2 B_2 +\gamma_{33} E_3 B_3 
      + \gamma_{32} E_3 B_2 + \gamma_{23} E_2 B_3 \rbrace \nonumber , 
  \end{eqnarray}
  from which the expressions for the nonzero components of the $\xi$, $\zeta$ 
  and $\gamma$ tensors in terms of Tresguerres' functions can be read for 
  the nonminimal torsion coupling considered here. Haugan \& Kauffmann \cite{hk95}
  have shown that the expressions ${\cal A}-{\cal C}$ and ${\cal B}$ from 
  (\ref{deltac}) can now be expressed in terms of the spherical components 
  of these tensors $\xi,\,\zeta$ and $\gamma$  
  \begin{equation}\label{exp.a-c}
    {\cal A}-{\cal C} = \frac{2}{\sqrt{6}}((\xi^{(2)}_{2'}+\xi^{(2)}_{-2'})
    +2i(\gamma^{(2)}_{2'}-\gamma^{(2)}_{-2'}) +(\zeta^{(2)}_{2'}+\zeta^{(2)}_{-2'}))
  \end{equation}
  and
  \begin{equation}\label{exp.b}
    {\cal B} = -\frac{1}{\sqrt{6}}(i(\xi^{(2)}_{2'}-\xi^{(2)}_{-2'})
    +2(\gamma^{(2)}_{2'}+\gamma^{(2)}_{-2'})+i(\zeta^{(2)}_{2'}-\zeta^{(2)}_{-2'})) \quad .
  \end{equation}
  Only $l=2$ components appear.  The $m'$ notation indicates that these 
  spherical components correspond to Cartesian components in a 
  quasi-Lorentzian $(t,x',y',z')$ coordinate system oriented 
  so that the radiation of interest propagates in the $z'$ direction.  
  In the following, we exploit the fact that these spherical 
  components can be expressed in terms of components in another 
  quasi-Lorentzian $(t,x,y,z)$ coordinate system via the familiar 
  transformation law \cite{edm}, e.g.
  \begin{equation}\label{transd}
    \xi^{(l)}_{m'}={\cal D}^{(l)}_{m'm}(\phi,\theta,\psi)\,\xi^{(l)}_m \quad ,
  \end{equation}
  where $\phi,\,\theta$ and $\psi$ are the Euler angles specifying the rotation 
  from $(t,x,y,z)$ to $(t,x',y',z')$ and the rotation matrix ${\cal D}$ is 
  given in terms of spherical harmonics. 
  
  Since the only direction that a spherical field can single out is the radial 
  one, it is now useful to introduce, at each point along a light ray of 
  interest, a local quasi-Lorentzian $(t,x,y,z)$ coordinate system oriented 
  so that the $z$ axis is radial and the $x$ axis lies in the ray's plane.  
  This is convenient because the spherical tensors introduced above are 
  simple in such local coordinate systems. Specifically, 
  $\xi^{(2)}_m,\,\zeta^{(2)}_m$ and $\gamma^{(2)}_m$ are nonzero only for $m=0$.
      
  At each point along the ray, the local $(t,x,y,z)$ coordinate 
  system is rotated about the $y$ axis through an angle $\theta$ to obtain a 
  local $(t,x',y',z')$ system so that now the ray runs in the $z'$ direction.  
  The local value of $\delta c/c$ in (\ref{deltac}) is expressed in terms of 
  ${\cal A}-{\cal C}$ and ${\cal B}$ which are, in turn, expressed in terms 
  of $\xi^{(2)}_{\pm 2'},\,\zeta^{(2)}_{\pm 2'}$ and $\gamma^{(2)}_{\pm 2'}$
  according to (\ref{exp.a-c}) and (\ref{exp.b}). Since the Euler angles 
  of the rotation from $(t,x,y,z)$ to $(t,x',y',z')$ are $\theta$ 
  and $\phi=\psi=0$, the transformation law (\ref{transd}) simplifies to 
  \begin{equation}
    \xi^{(2)}_{\pm 2'}=\sin^2\theta\,\xi_0^{(2)} \quad ,
  \end{equation}
  with the same relationship between $\zeta^{(2)}_{\pm 2'}$ and $\zeta^{(2)}_0$ 
  and between $\gamma^{(2)}_{\pm 2'}$ and $\gamma^{(2)}_0$. Here, $\theta$ 
  denotes the angle between the light ray's propagation direction and the 
  radial direction. Exploiting these transformations, Haugan \& Kauffmann 
  conclude from (\ref{exp.a-c}) and (\ref{exp.b}) that 
  \begin{equation}
    {\cal A}-{\cal C}=\frac{4}{\sqrt{6}}(\xi_0^{(2)}+\zeta_0^{(2)})\sin^2\theta 
  \end{equation}
  and  
  \begin{equation}
    {\cal B} = -\frac{4}{\sqrt{6}}\left(\gamma^{(2)}_0\sin\theta\right)^2 \quad.
  \end{equation}
  Substitution into (\ref{deltac}) yields an expression for the fractional 
  difference in the propagation speeds of linear polarization states for 
  the case of a static spherically symmetric torsion field
  \begin{equation}\label{deltac2}
    \frac{\delta c}{c} = \sqrt{\frac{2}{3}}\,\sin^2\theta\,\sqrt{\left(
    \xi_0^{(2)}+\zeta_0^{(2)}\right)^2+\left(4\gamma^{(2)}_0\right)^2} \quad .
  \end{equation}    
  To express this in terms of Tresguerres' parameters we simply use 
  the relationship between the spherical and Cartesian components of 
  $\xi$, $\zeta$ and $\gamma$ and between those Cartesian components 
  and Tresguerres' parameters.  
  \begin{eqnarray*}
    \xi_0^{(0)} &=& \xi^{11}+\xi^{22}+\xi^{33} = 2k^2\,(\gamma_{(3)}^2+\gamma_{(4)}^2) \\
    \xi_0^{(2)} &=& \xi^{11}-\frac{1}{2}(\xi^{22}+\xi^{33})=-k^2(\gamma_{(3)}^2
                    +\gamma_{(4)}^2)\\
    \zeta_0^{(0)} &=& \zeta^{11}+\zeta^{22}+\zeta^{33}=k^2[(\alpha^2-\beta^2)-2 
                    (\gamma_{(1)}^2+\gamma_{(2)}^2)] \\
    \zeta_0^{(2)} &=& \zeta^{11}-\frac{1}{2}(\zeta^{22}+\zeta^{33}) = k^2[(\alpha^2 - \beta^2) 
                      + (\gamma_{(1)}^2+\gamma_{(2)}^2) ]\\
    \gamma^{(0)}_0 &=& \gamma^{11}+\gamma^{22}+\gamma^{33}=2k^2\,(\gamma_{(1)}\gamma_{(3)}-
                       \gamma_{(2)}\gamma_{(4)})\\
    \gamma^{(2)}_0 &=& \gamma^{11}-\frac{1}{2}(\gamma^{22}+\gamma^{33})
                       =-k^2(\gamma_{(1)}\gamma_{(4)} -\gamma_{(2)}\gamma_{(3)}) 	\quad .	           	     
  \end{eqnarray*}

  For the particular torsion field (\ref{t_sol}) we have 
  $\gamma_{(2)}=\gamma_{(4)}=0$ and, 
  therefore, ${\cal B}=0$. Using (\ref{t_sol}) one gets
  \begin{eqnarray}
    \xi_0^{(2)} &=& -\frac{k^2}{4}\left(\frac{1}{r^2}-\frac{2m}{r^3}+\frac{m^2}{r^4}\right) \\
    \zeta_0^{(2)} &=& -\frac{4k^2m}{r^3}+\frac{k^2}{4}\left(\frac{1}{r^2}+\frac{2m}{r^3}+\frac{m^2}{r^4}\right)
  \end{eqnarray} 
  which yields
  \begin{equation}
    \xi_0^{(2)} + \zeta_0^{(2)} = -\frac{3k^2m}{r^3} \quad .
  \end{equation}
  so that we have
  \begin{equation}
    {\cal A}-{\cal C}=-\frac{12\,k^2\,m}{\sqrt{6}\,r^3}\sin^2\theta \quad .
  \end{equation}
  Therefore, the fractional difference between the velocities of the two polarization states 
  is given by
  \begin{equation}
    \frac{\delta c}{c} = -\sqrt{6}\frac{k^2m}{r^3}\sin^2\theta \quad ,
  \end{equation}
  where $\theta$ again denotes the angle relative to the outward radial direction in
  which the light is propagating. For the total phase shift $\Delta\Phi$ which 
  accumulates between the source and the observer one now has to calculate
  \begin{equation}
    \omega\int\frac{\delta c}{c}\,dt = -\sqrt{6}\omega k^2m\int\frac{\sin^2\theta}{r^3}\, dt \quad .
  \end{equation}
  Recapitulating the analysis performed in \cite{gab91,gab91a}, we find
  \begin{equation}
    \Delta \Phi = \sqrt{2 \over 3} {{2 \pi k^2 m} \over {\lambda R^2}} 
    {{(\mu +2) (\mu - 1)} \over {\mu + 1}}, 
    \label{mag-phase}
  \end{equation}
  where $\mu$ denotes the cosine of the light source's heliocentric angle, 
  $\lambda$ is the light's wavelength and $R$ is the Sun's radius.  
\section{Technique}
  We follow two strategies to test for gravitational birefringence. One of these
  was outlined by Solanki \& Haugan (1996)\cite{sh96}, but could not be applied due 
  to a lack of appropiate data. It is summarized and its 
  implementation is described in Sec. III.B. The other technique is new and is 
  described in Sec. III.A. In order to compare the effectiveness of these techniques
  with results of previous attempts to set limits on gravity-induced birefringence,
  which dealt with NGT, we also briefly consider NGT here in addition to metric-affine theories
  of gravity.
  \subsection{Profile Difference Technique}
    This technique relies on the fact that $\Delta\Phi$ is expected to be a strong 
    function of $\mu$, which is confirmed in the two concrete cases of NGT (see reference
    \cite{gab91b}) and metric-affine theories (see Sect. II). $\Delta\Phi$ is the phase 
    shift which accumulates as light propagates from a point on the solar surface to 
    the observer between net circular polarization, described by the Stokes parameter 
    $V$, and net linear polarization aligned at $45^{\circ}$ to the nearest part of the
    solar limb, generally ascribed to Stokes $U$. Formulae for $\Delta\Phi$ as predicted by
    metric-affine theories are given in Sect.2. For Moffat's NGT (Moffat 1979 \cite{mof79}) 
    a corresponding expression has been published by Gabriel et al. (1991)\cite{gab91,gab91b}.     
    This means that for any sufficiently large NGT charge $\ell_{\odot}$ or equivalent 
    metric-affine parameter $k$ a mixture of Stokes $V$ and $U$ profiles will be 
    observed from sources of polarization distributed across the solar disc (i.e. as a
    function of $\mu$), irrespective of the exact polarization state of the emitted radiation
    (i.e. which mixture of $V$ and $U$ is produced at the solar surface). Let the subscripts 
    'src' and 'obs' signify the Stokes profiles as created at the source and as observed, 
    respectively. Then, irrespective of the value of $<|V_{\src}|>-<|U_{\src}|>$ for 
    sufficiently large $\ell_{\odot}$ or $k$, $<|V_{\obs}|>-<|U_{\obs}|>$ will tend 
    to zero. The averaging is over all $\mu$ values and the total number of profiles is 
    assumed to be large.
    
\begin{figure}         
     \centerline{\resizebox{2.1in}{2.1in}{\includegraphics[angle=90]{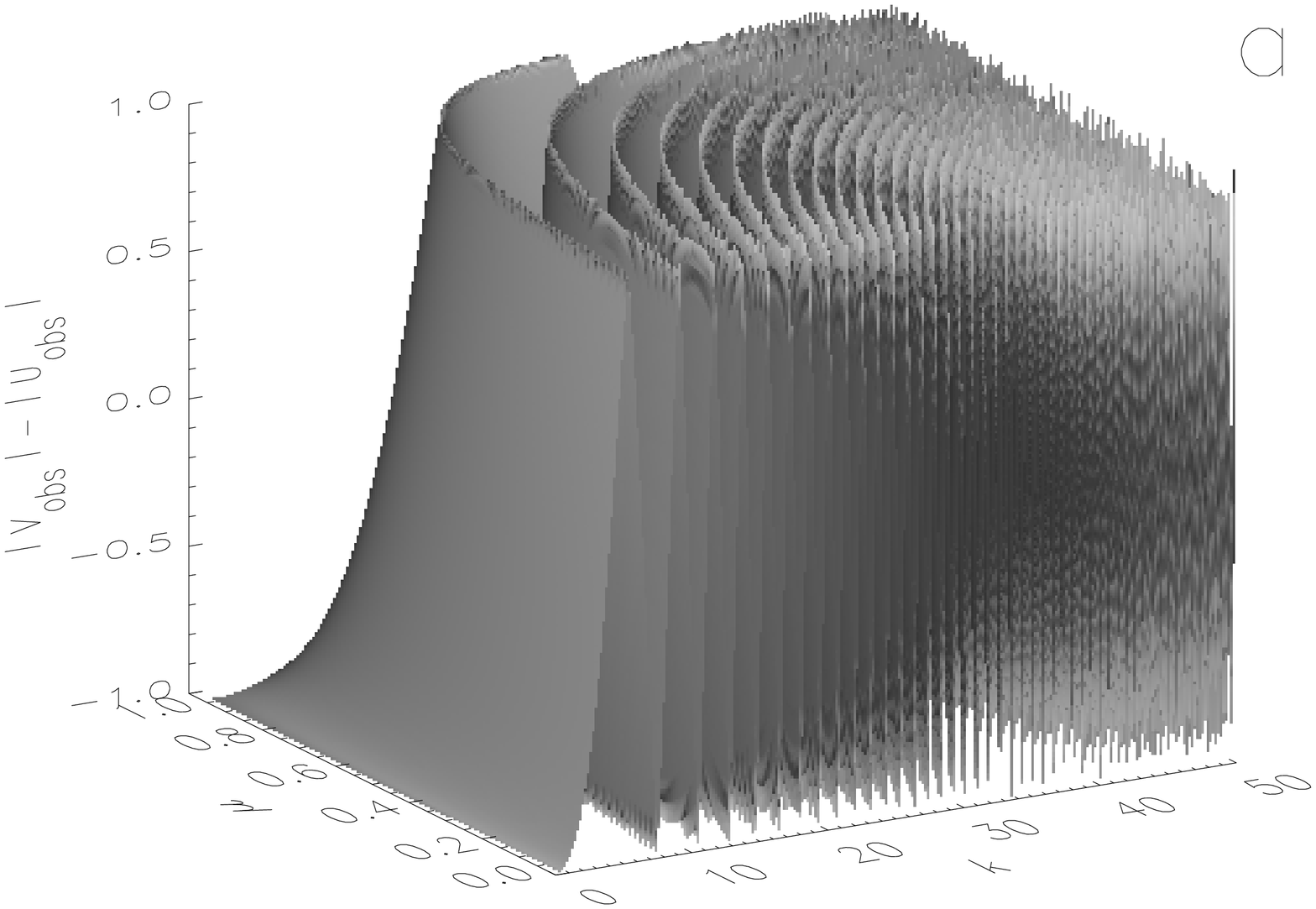}}
                 \resizebox{2.1in}{2.1in}{\includegraphics[angle=90]{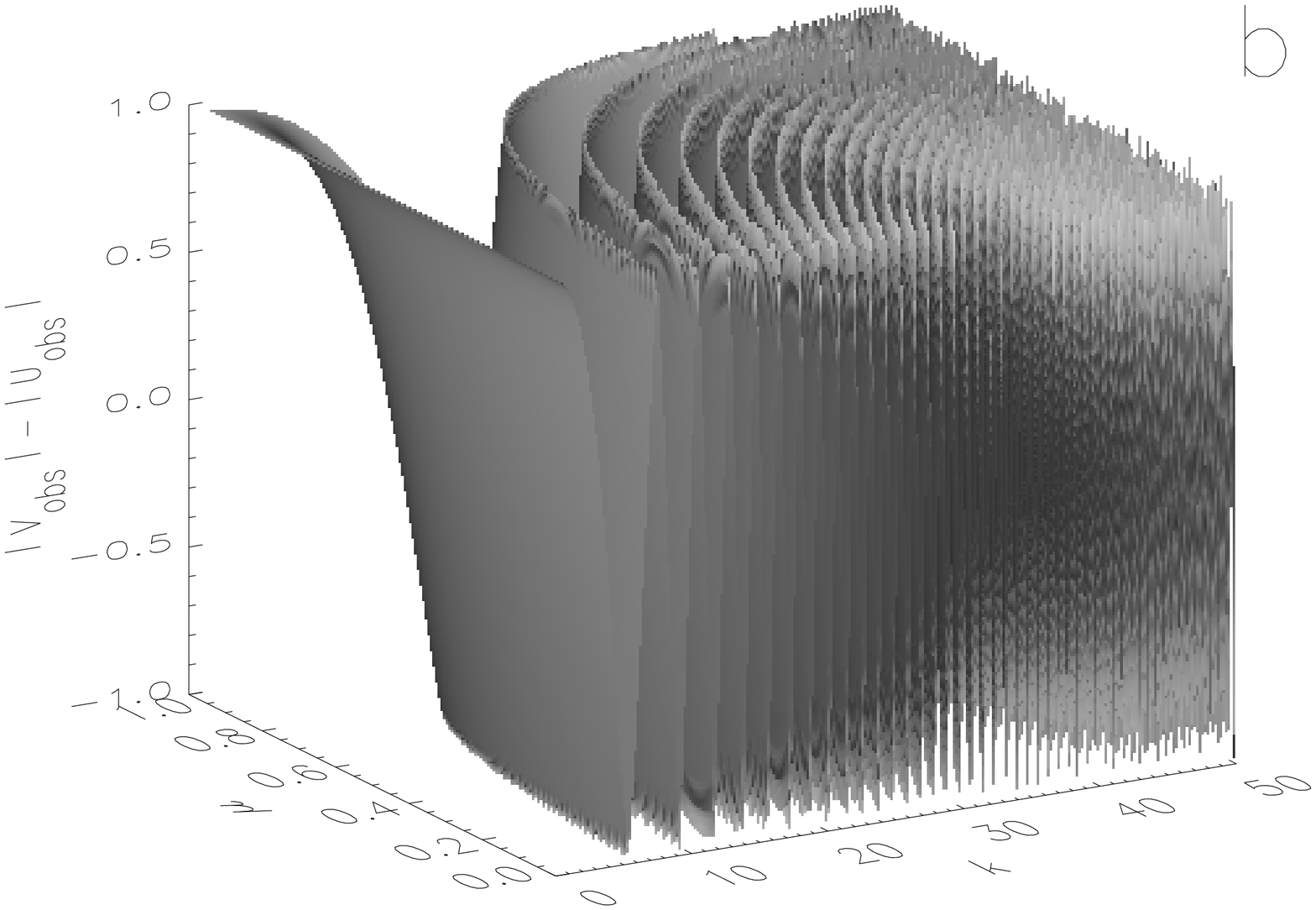}}}
     \centerline{\resizebox{2.1in}{2.1in}{\includegraphics[angle=90]{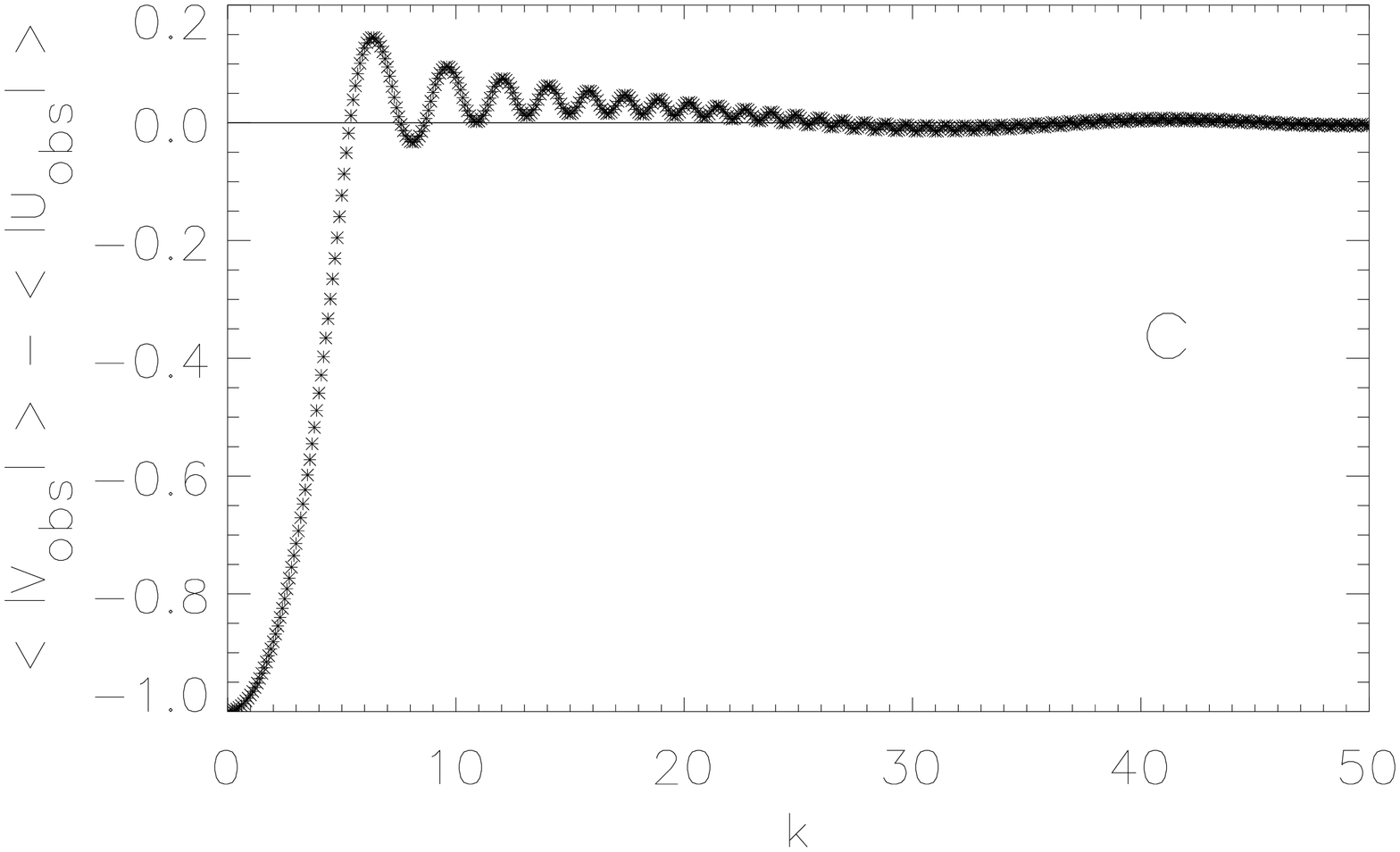}}
                 \resizebox{2.1in}{2.1in}{\includegraphics[angle=90]{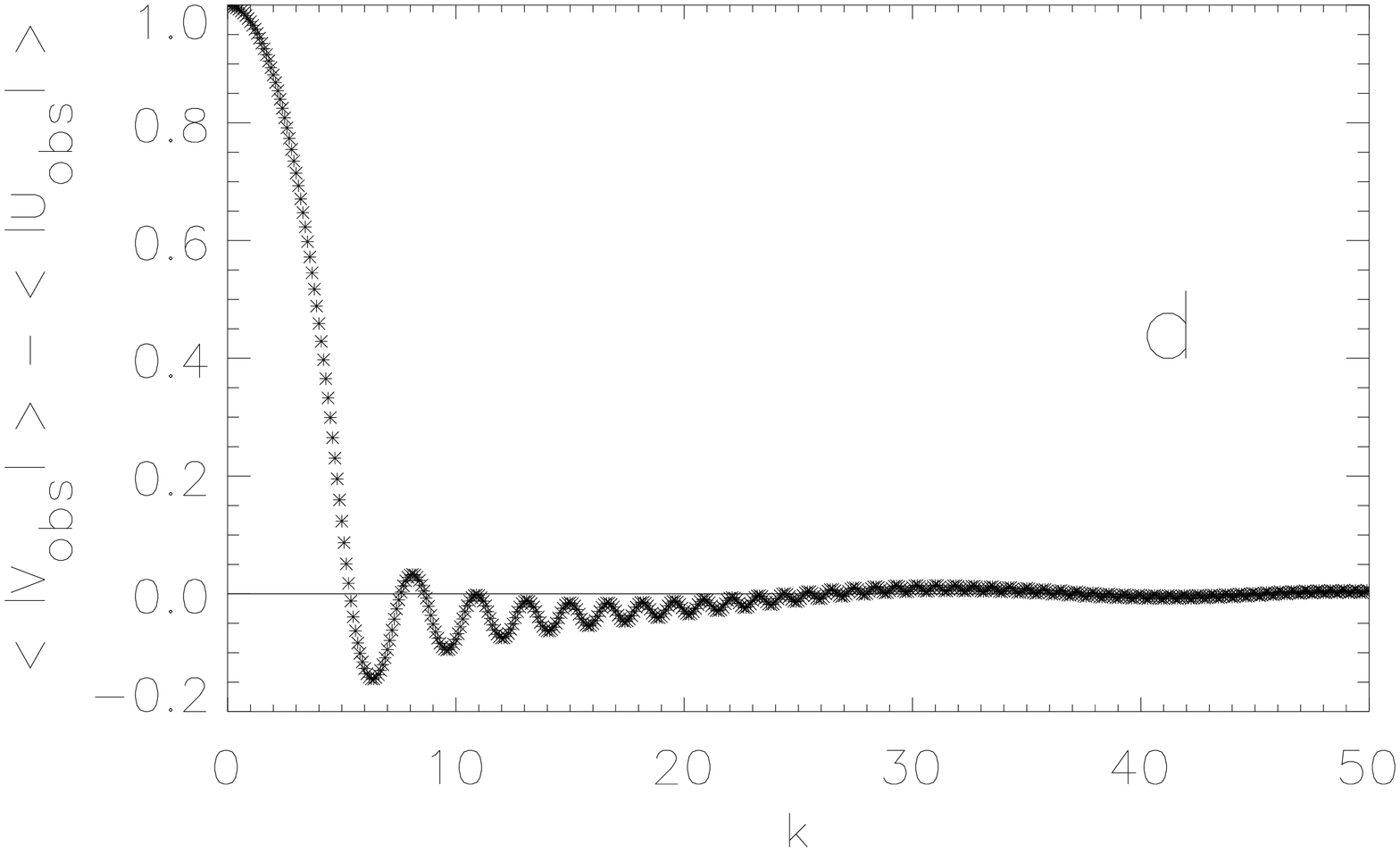}}}
      \caption{Top: $\mid V_{\obs}| - |U_{\obs}|$ vs. $k$ and $\mu$ for 
               $|V_{\src}| = 0$ (a) and $|U_{\src}| = 0$ (b).
	       Bottom: $|V_{\obs}| - |U_{\obs}|$ averaged over all $\mu$ for 
	       the above cases. }
      \label{pdt1}
    \end{figure}

    This effect is illustrated in Fig.~\ref{pdt1}. In Figs.~\ref{pdt1}a \& b 
    $|V_{\obs}| - |U_{\obs}|$ is plotted vs. $k$ and $\mu$ for the extreme 
    cases $|V_{\src}| = 0$ (Fig.~\ref{pdt1}a) and $|U_{\src}| = 0$ 
    (Fig.~\ref{pdt1}b). Other combinations of $|V_{\src}(\mu)|$
    and $|U_{\src}(\mu)|$ give qualitatively similar results.
    The $|V_{\obs}| - |U_{\obs}|$ surface oscillates ever more rapidly with increasing
    $k$ and with decreasing $\mu$. Lines of equal $|V_{\obs}| - |U_{\obs}|$
    are strongly curved in the $k - \mu$ plane. These two points combine to lead
    to decreasing $<|V_{\obs}|>-<|U_{\obs}|>$ with increasing $k$. This is shown 
    in Figs.~\ref{pdt1}c \& d for the cases illustrated in Figs. ~\ref{pdt1}a and b,
    respectively. As expected, the $<|V_{\obs}|>-<|U_{\obs}|>$ vs. $k$ curves exhibit a rapidly 
    damped oscillation around zero. This effect can be used to set upper limits on 
    gravitational birefringence if the observations exhibit a $ <|V_{\obs}|>-<|U_{\obs}|>$ 
    that differs significantly from zero. As we show later in this paper, this 
    is indeed the case.

  \subsection{Stokes asymmetry technique}   
    The strategy proposed by Solanki \& Haugan (1996)\cite{sh96} makes use of the symmetry 
    properties of the Stokes profiles produced by the Zeeman splitting of atomic 
    spectral lines. In the absence of radiative transfer effects in a strongly dynamic 
    medium net circular polarization, Stokes $V$, is antisymmetric in wavelength 
    and net linear polarization aligned at $45^{\circ}$ to the solar limb, Stokes $U$, 
    is symmetric. Birefringence, predicted by metric-affine gravity theories (and NGT) 
    changes the phase between orthogonal linear polarizations and thus partly converts 
    Stokes $V$ into Stokes $U$ and vice versa. However, $U$ produced from $V$ by 
    gravitational birefringence still has the symmetry of $V$ and can thus be 
    distinguished from the Zeeman signal.
  
    Let subscripts '$a$' and '$s$' signify the antisymmetric and symmetric parts of the
    Stokes profiles, respectively. Then,
    \begin{eqnarray}
      \frac{U_{\asc}}{U_{\ssc}} &=& 
      \frac{V_{\aob}\sin\Delta\Phi + U_{\aob}\cos\Delta\Phi}
           {V_{\sob}\sin\Delta\Phi + U_{\sob}\cos\Delta\Phi}  \label{U_{src}} \quad , \\
      \frac{V_{\ssc}}{V_{\asc}} &=& 
      \frac{V_{\sob}\cos\Delta\Phi + U_{\sob}\sin\Delta\Phi}
           {V_{\aob}\cos\Delta\Phi + U_{\aob}\sin\Delta\Phi} \label{V_{src}} \quad .
    \end{eqnarray}
    Thus for observed symmetric and antisymmetric fractions of $U$ and $V$ Eqs. (3.1)
    and (\ref{V_{src}}) predict the ratios $U_{\fa}/U_{\fs}$ and 
    $V_{\fs}/V_{\fa}$ at the solar source.
  
    If the solar atmosphere were static these ratios would vanish $(U_{\asc}=V_{\asc}=0)$, 
    so that any observed $U_{\fa}$ or $V_{\fs}$ would be due to either $\Delta\Phi$ 
    or noise: $U_{\aob} = V_{\asc} \sin\Delta\Phi$, $V_{\sob} = U_{\ssc}\sin\Delta\Phi$. 
    The solar atmosphere is not static, however, and consequently the Stokes profiles do 
    not fulfill the symmetry properties expected from the Zeeman effect even for rays 
    coming from solar disc centre $(\mu = 1)$, which are unaffected by gravitational birefringence. 
    This asymmetry has been extensively studied, in particular for Stokes $V$, which 
    most prominently exhibits it \cite{sos84,gd89,sgs99,mp97}. Although most profiles 
    have $V_{\fs}/V_{\fa} \lsim 0.2$, a few percent of $V$ profiles exhibit $V_{\fs}/V_{\fa}$ 
    values close to unity, even at $\mu = 1$. Such profiles occur in different 
    types of solar regions, e.g. the quiet Sun \cite{sgs99}, active region 
    neutral lines \cite{sea93} and sunspots \cite{sal92}. 
    The magnitude of $V_{\fs}/V_{\fa}$ decreases rapidly with increasing $V_{\fa}$ and 
    profiles with $V_{\fs}/V_{\fa} \gsim 1$ are all very weak. They are often associated 
    with the presence of opposite magnetic polarities within the spatial resolution element 
    and a magnetic vector that is almost perpendicular to the line of sight, situations
    which naturally give rise to small $V$ \cite{sal92,plo01}.     
  
    The observed Stokes $U$ asymmetry is on average smaller than the $V$ asymmetry. This is 
    true in particular for extreme asymmertic values, i.e. \\$(U_{\aob}/U_{\sob})_{\mbox{\footnotesize{max}}} 
    \ll (V_{\sob} /V_{\aob})_{\mbox{\footnotesize{max}}}$. Since this relation also holds at 
    $\mu = 1$ it is valid for source profiles as well. Thus, S\'{a}nchez Almeida \& Lites 
    (1992) \cite{sal92} point out that Stokes $U$ retains $U_{\aob}/U_{\sob} \ll 1$ throughout 
    a sunspot, although $V_{\fs}/V_{\fa} > 1$ is invariably achieved at the neutral line. 
    The reason for the smaller maximum asymmetry lies in the fact that Stokes $U$ senses the 
    transverse magnetic field. Since velocities in the solar atmosphere are directed mainly 
    along the field lines they generally have a small line-of-sight component when $U$ has a
    significant amplitude. Sizable line-of-sight velocities are needed, however, to produce a
    significant asymmetry \cite{gd89}. Another reason for the smaller maximum $U$ asymmetry 
    is that, unlike Stokes $V$, it does not distinguish between oppositely directed magnetic fields.
  
    Thus it is not surprising that in the following analysis Stokes $U$ provides tighter 
    limits than Stokes $V$. Another reason is that due to the on average stronger observed $V$
    profiles, asymmetries introduced in $U$ (through gravitationally introduced cross-talk from $V$)
    are larger than the other way round. However, we also analyse Stokes $V$ as a consistency
    check. 

    In order to seperate the asymmetry produced by solar effects from that introduced 
    by gravitational birefringence, one strategy to follow is to consider large amplitude
    Stokes profiles only. Another is to analyse data spanning a large range of $\mu$ values, 
    since $\Delta\Phi$ exhibits a definite $\mu$ dependence. Finally, the larger the 
    number of analysed line profiles, the more precise the limit that can be set on 
    $\Delta\Phi$. Better statistics not only reduce the influence of noise, they are 
    also needed because for a single profile gravitational birefringence can both 
    increase or decrease $V_{\fs}/V_{\fa}$ and $U_{\fa}/U_{\fs}$. The latter may become 
    important if the source profiles are strongly asymmetric. Thus a small observed 
    $V_{\fs}/V_{\fa}$ or $U_{\fa}/U_{\fs}$ is in itself no guarantee for a small 
    gravitational birefringence. However, since almost all source profiles are 
    expected to have $V_{\fs}/V_{\fa} \ll 1$, $U_{\fa}/U_{\fs} \ll 1$, on average 
    we expect gravitational birefringence to increase these ratios. 

\section{Observations and data}
  Two sets of data have been analysed in the present paper. They are described
  below.
  \subsection{Data obtained in 1995}
    Observations were carried out from 7 - 13th Nov. 1995 with the Gregory 
    Coud\a'e Telescope (GCT) at the Teide Observatory on the
    Island of Teneriffe. For the polarimetry we employed the first version 
    of the Z\"urich Imaging Polarimeter (ZIMPOL I), which employs 3 CCD cameras,
    one each to record Stokes $I\pm Q,$ $I\pm U$ and $I\pm V$ simultaneously 
    \cite{pov95}. 
    
    The recorded wavelength range contains four prominent spectral lines, 
    Fe I 5247.06{\AA}, Cr I 5247.56{\AA}, Fe I 5250.22{\AA} \& Fe I 5250.65{\AA}.  
    Three of these spectral lines are among those with the largest Stokes amplitudes 
    in the whole solar spectrum and are also unblended by other spectral lines  
    \cite{sea86}. Blending poses a potentially serious problem since it can affect 
    the blue-red asymmetry of the Stokes profiles. By analysing more than one such line it 
    is possible to reduce the influence of hidden blends and noise.
    Nowhere else in the visible spectrum are similar lines located sufficiently
    close in wavelength that they can be recorded simultaneously on a single detector. 
    Also, compared to other lines with large Stokes amplitudes the chosen set lies 
    at a short wavelength. This enhances $\Delta\Phi$ since it scales with 
    $1/\lambda$. The sum of the above properties make the chosen range uniquely 
    suited for our purpose.

    In order to image all 4 spectral lines of interest onto a single CCD we introduced
    reduction optics between the image plane of the spectrograph and the detectors.
    They produce an image-scale reduction by a factor of 3.2. The final spectral 
    resolving power $\lambda/\Delta\lambda$ corresponded to 210'000. The spatial 
    scale corresponding to a pixel was $1.13''$ (or 860 km on the Sun). However, 
    the effective spatial resolution of the data is limited by turbulence in the Earth's 
    atmosphere, so-called 'seeing'. This varied somewhat in the course of the 
    observing run, so that the estimated angular resolution of the observations 
    lies between $2.2''$ and $3''$.
    
    The modulator package, composed of 2 photoelastic modulators oscillating
    at frequencies of around 41 kHz \& 42 kHz followed by a Glan linear polarizer,
    was placed ahead of the entrance slit to the spectrograph, but was nevertheless
    (unavoidably) located after 2 oblique reflections in the telescope. Oblique 
    reflections produce cross-talk between Stokes parameters, i.e. they partially
    convert one form of polarization into another. Since we are trying to observe, 
    or at least set limits on ''cross-talk'' between Stokes $U$ and $V$ due to 
    gravitational birefringence we took some trouble to reduce the instrumental 
    cross-talk to the extent possible. A first step was the choice of the telescope. 
    With only two oblique reflections, whose relative angles change only slowly in
    the course of a year, the GCT is relatively benign compared to most other large 
    solar telescopes. Secondly, a half-wave plate was introduced between the two 
    oblique reflections. S\'{a}nchez Almeida et al. \cite{smw91,smk95} have 
    pointed out that a half-wave plate at that location should, under ideal circumstances, 
    completely eliminate all instrumental cross-talk. To test the efficiency of the
    half-wave plate in suppressing instrumental cross-talk between Stokes $Q$, $U$ and $V$ 
    we first carried out a series of observations of a sunspot near the centre of the 
    solar disc both with and without a half-wave plate introduced in the light path. 
    Such tests were necessary since the half-wave plate available at the GCT is not
    optimized for the observed wavelength. Note that at solar disc centre $(\mu = 1)$ 
    $\Delta\Phi=0$, so that we test for instrumental cross-talk 
    only. The half-wave plate was indeed found to significantly reduce instrumental 
    cross-talk. Remaining cross-talk was removed during data reduction using a numerical 
    model of the telescope that includes an imperfect half-wave plate (adapted 
    from a model kindly provided by V. Mart\'{\i}nez Pillet). The parameters of the 
    model were adjusted slightly using the observations of a sunspot umbra located close to 
    $\mu = 1$. We estimate that the residual cross-talk after this procedure is 
    at the level of a few percent. Since Stokes $V$, $Q$, $U$ profiles generally have 
    amplitudes of $0.1I_c$ or less, the influence of the cross-talk is of the same order as
    the noise, which is roughly $1-2\times10^{-3} I_c$, where $I_c$ is the continuum 
    intensity. Photon noise is by far the largest contributor to this noise level. 
    At this level instrumental cross-talk ceases to be of concern for our analysis.
    
    The Zimpol polarimeters are unique in that they combine CCD detectors with a very
    high modulation frequency $(\geq 40 $ kHz) and hence preclude distortion of the
    Stokes profiles and cross-talk between them due to seeing fluctuations. The other 
    advantage of the ZIMPOL working principle is the fact that the fractional polarization
    is absolutely free from gain-table or flat field noise \cite{pov95}. This again 
    improves the accuracy of the profile shapes and hence the reliability of our results.
    
    A total of 106 recordings were made at different locations on the solar disc in 
    an attempt to cover a large range of $\mu$ homogenously. Particular emphasis was 
    placed on observations close to the limb since gravitational birefringence is expected 
    to be largest for such rays.
     
    Since only a single sunspot was present on the solar disc during the observing run 
    most recordings refer to faculae and network features, i.e. magnetic features with
    lower Stokes $Q$, $U$, $V$ signals. 
    \begin{figure} 
      \vspace{0.5cm}
      \resizebox{4.1in}{6.65in}{\includegraphics{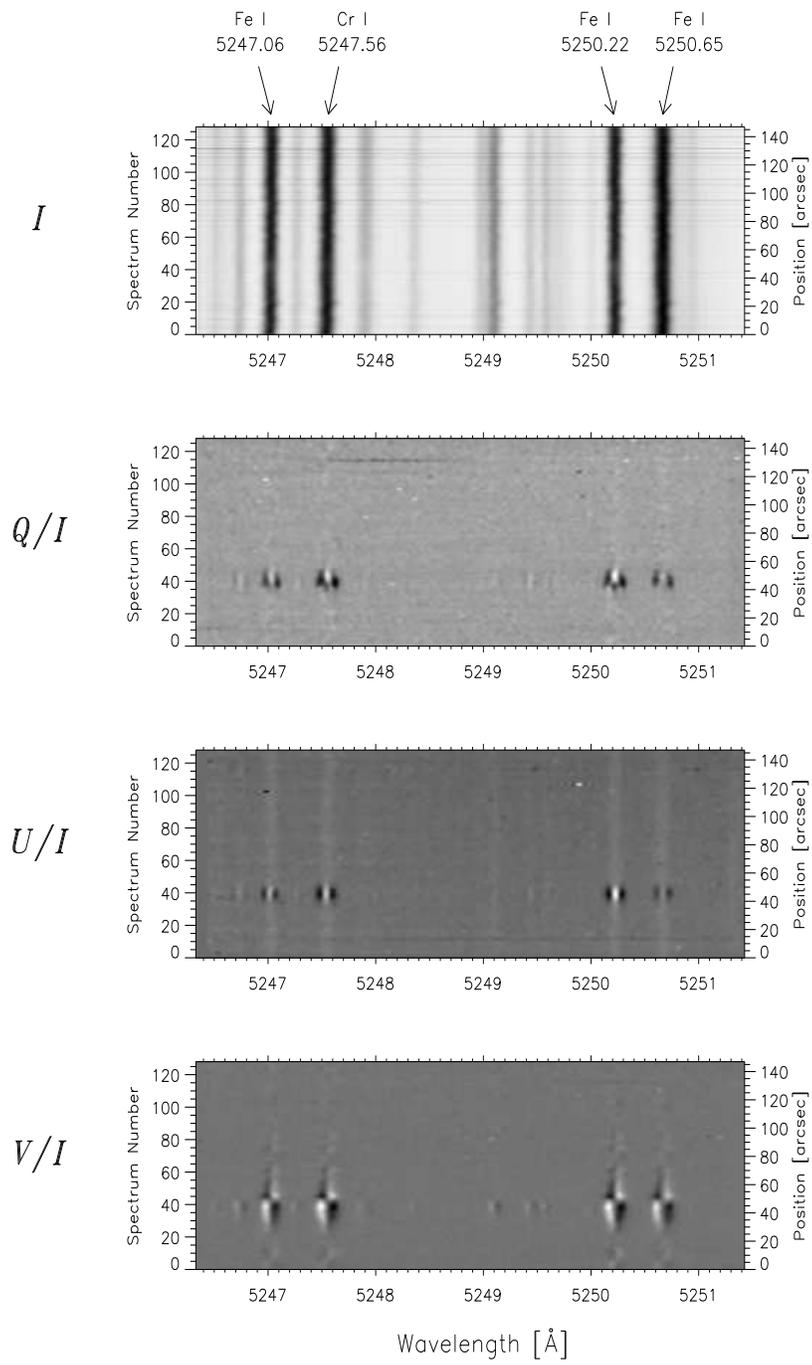}}      
      \vspace{0.5cm}
      \caption{Sample spectrum of Stokes $I, Q, U$ and $V$.}
      \label{n4fac}
    \end{figure}       
    In Fig.~\ref{n4fac} a sample Stokes $I,Q,U,V$ spectrum of a facular region near the solar 
    limb is plotted. The 4 analyzed spectral lines are identified. These data were fully 
    reduced following the tedious, but well-tested procedures described by Bernasconi (1997)
    \cite{ber97}.
    
  \subsection{Data set of March 2000}
    In order to improve the statistics and the $\mu$ coverage a second observing run was 
    carried out in March 2000 with the Gregory - Coud\a'e Telescope of IRSOL (Istituto
    Ricerche Solari Locarno) in Locarno, Switzerland. 
    This telescope is almost identical to the GCT on Tenerife and the parameters such as 
    spectral resolution, noise level etc. are very similar to those of the 1995 observations.
    
    The next generation, ZIMPOL II polarimeter \cite{gap97,pov98} was employed
    for the polarization analysis and data recording. It simultaneously records three
    of the four Stokes parameters, either Stokes $I,Q,V$ or $I,U,V$ on a single CCD detector 
    chip. Observations in these two modes were interlaced, such that alternate
    exposures record Stokes $I,Q,V$ and $I,U,V$, respectively.
    Exposures of the same Stokes parameters were then added together to reduce noise.
    Thus the final data set consists of all four Stokes parameters. The only differences 
    with respect to the recordings made in 1995 are that the number of 
    spatial pixels is reduced and that the noise level of Stokes $V$ in the newer 
    data is a factor of $\sqrt{2}$ lower than of Stokes $Q$ and $U$, whereas Stokes 
    $Q$, $U$ and $V$ had the same noise level in the earlier recordings. Due to the 
    superior modulation scheme implemented in ZIMPOL II Stokes $Q$ and
    $U$ achieve a noise level of $10^{-3}I_c$ after roughly the same exposure time 
    during the observations made in 2000 as during the earlier campaign.
    
    These observations were carried out on the day of the equinox, at which time
    the two mirrors producing oblique reflections of the beam ahead of the modulator package
    are oriented such that their polarization cross-talk cancels out. Hence for
    these observations the instrumental cross-talk is essentially zero and no further
    treatment of the data for this effect is required.

    The Sun was very active at the time of these observations with many active regions
    harboring sunspots and faculae present on the solar disc. Since active regions
    generally give larger amplitude Stokes signals we concentrated on observing them. 
    A total of 7 exposures were made. The typical seeing during these observations was estimated
    to be 2 - $3''$, while the spatial pixel size was $1.13''$.

\section{Data analysis}
  Each exposure gives us the profiles of the 4 analyzed spectral lines in Stokes $I, Q, U$
  and $V$ at a set of 94 (128 in the 1995 data) positions on the solar disc. Once the
  reduction and calibration procedure is completed we select from a given frame those 
  spectra for further analysis for which the $S/N$ ratio for either Stokes $U$ or $V$ 
  in at least one of the four spectral lines is above 12 in the 1995 data and above 15 
  in the 2000 data.

  This criterion gave us a total of 4480 profiles for further analysis. A non-orthogonal 
  wavelet-packets smoothing scheme was employed to enhance the $S/N$ ratio by a factor of 
  1.5 - 2 without significantly affecting the profile shapes \cite{fas98}. 
  
\begin{figure}[t]     
    \centerline{\resizebox{2.4in}{2.4in}{\includegraphics{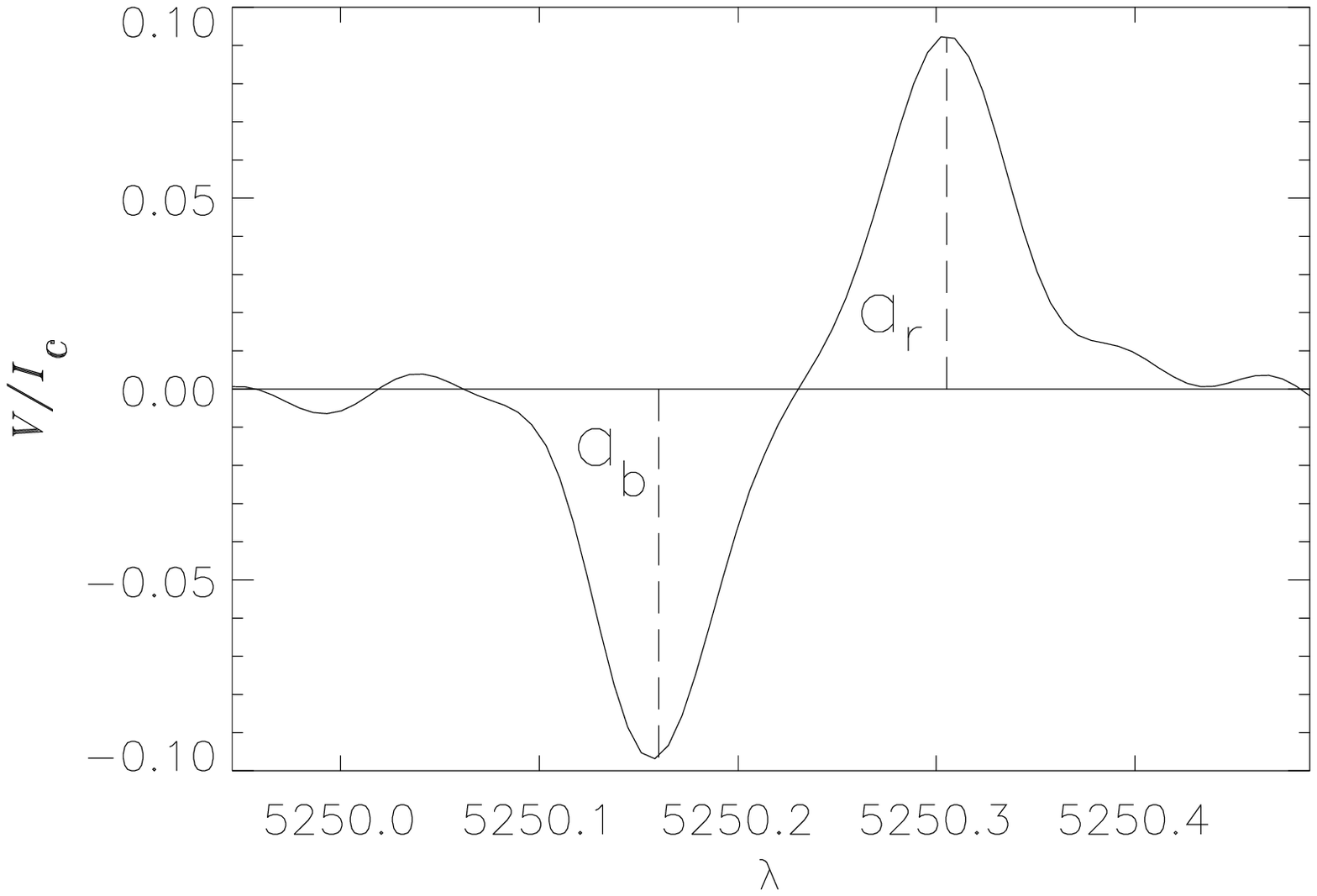}}
                \resizebox{2.4in}{2.4in}{\includegraphics{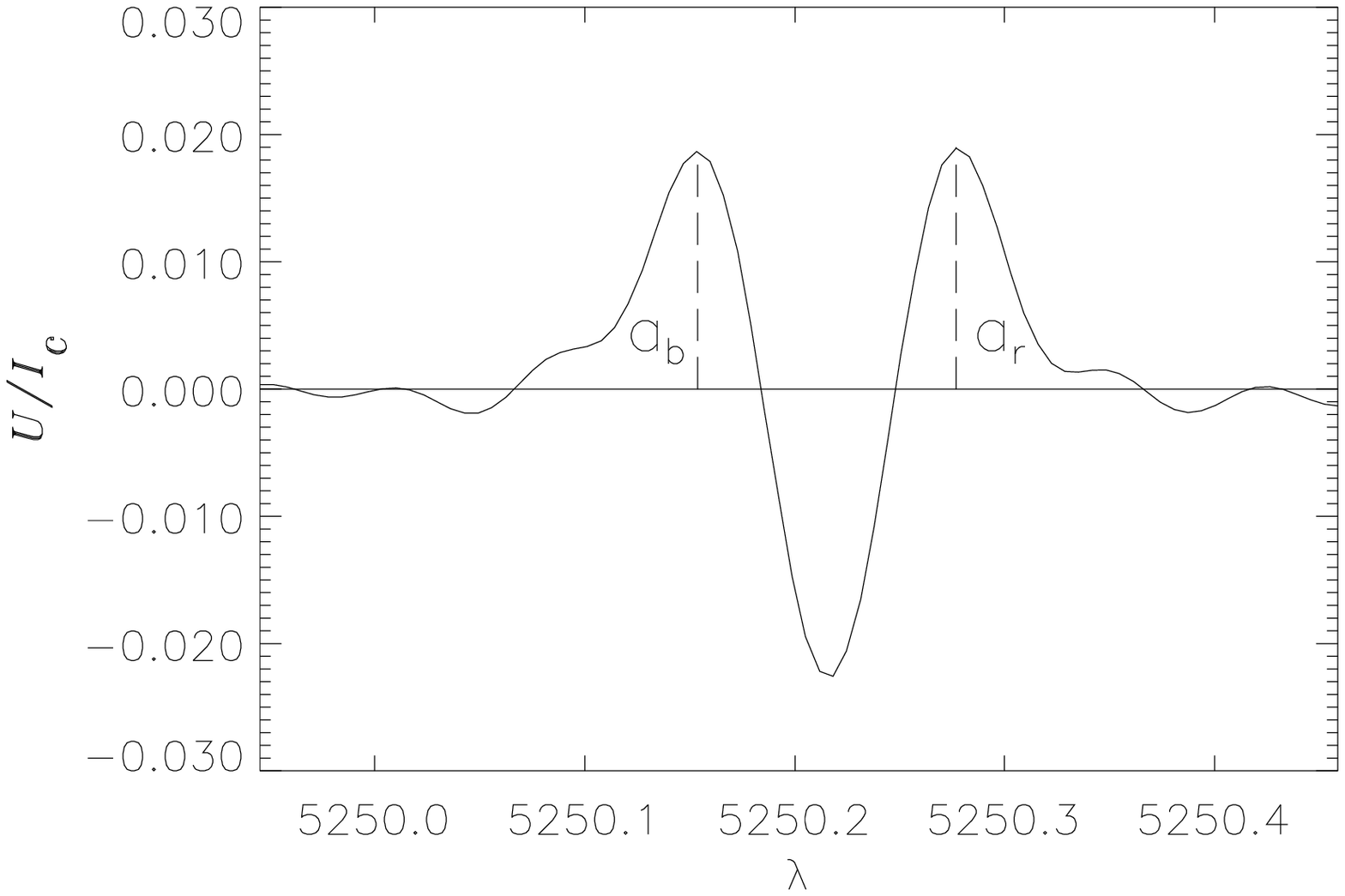}}}
    \vspace{0.27cm}			
    \caption{Simultaneously measured profiles of Stokes $V$ and Stokes $U$
             of the Fe I line at 5250.22{\AA}. }
    \label{fig2}
  \end{figure}

  Then the signed amplitudes of the blue and red wings, $a_{\mbox{\footnotesize{b}}}$ and 
  $a_{\mbox{\footnotesize{r}}}$ (i.e. of the blue and red Zeeman $\sigma$-component) of all 
  Stokes profiles of all 4 spectral lines was determined. In Fig. \ref{fig2} we plot 
  observed Stokes $U$ and $V$ profiles of the Fe I line at 5250.22{\AA}. 
  $a_{\mbox{\footnotesize{b,r}}}(V)$ and $a_{\mbox{\footnotesize{b,r}}}(U)$ are 
  indicated in the figure. 
  Using these we can form the symmetric and antisymmetric parts of the Stokes $V$ and $U$ 
  profile amplitudes, $V_{\fs}=(a_{\mbox{\footnotesize{b}}} + a_{\mbox{\footnotesize{r}}})/2$,
  $V_{\fa}=(a_{\mbox{\footnotesize{b}}} - 
  a_{\mbox{\footnotesize{r}}})/2$, $U_{\fs}=(a_{\mbox{\footnotesize{b}}} + 
  a_{\mbox{\footnotesize{r}}})/2$ and $U_{\fa}=(a_{\mbox{\footnotesize{b}}} - 
  a_{\mbox{\footnotesize{r}}})/2$, respectively, which enter Eqs. (\ref{U_{src}}) 
  and (\ref{V_{src}}). In the following all $V$ and $U$ values and parameters are normalized 
  to the continuum intensity, even when not explicitely mentioned.
  \subsection{Profile difference analysis}
\begin{figure}[t]
    \vspace*{-0.3cm}    
      \resizebox{4.in}{3.2in}{\includegraphics[angle=90]{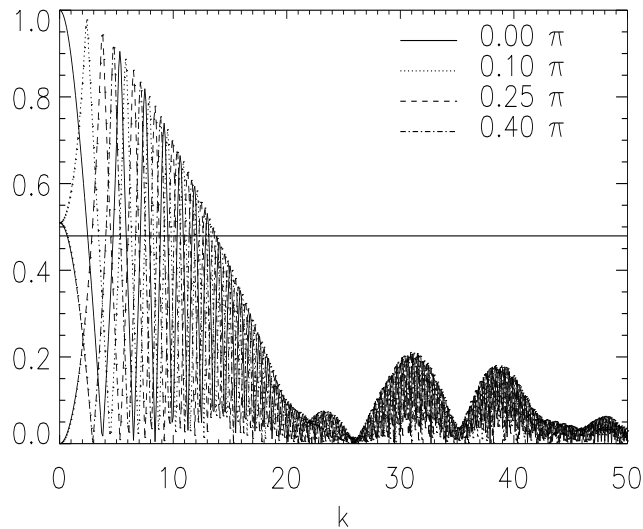}}
      \caption{The observable (Stokes $V$, Stokes $U$) mixture for initial phase
               differences of $0\pi$, $0.1\pi$, $0.25\pi$ and $0.4\pi$ plotted vs.
	       MAG coupling constant $k$. The horizontal solid line represents 
	       the value obtained from observations. Note that $\Delta\Phi$ 
	       values bigger than $0.5\pi$ give cyclic results.}
      \label{pdtl3}
    \end{figure}

    We now apply the technique outlined in Sec.III.A to our data. Due to the limited 
    number of profiles and their irregular distribution over $\mu$ (see Sect. V. B)
    any limit on gravitational birefringence will be less tight than what is achievable
    with ideal data presented in Sect. III.A. 
    In Fig.~\ref{pdtl3} we plot 
    \begin{equation}
      \frac{|<|V_{\obs}|>-<|U_{\obs}|>|}{<|V_{\obs}|>+<|U_{\obs}|>} 
      \quad \mbox{vs.} \quad k,
    \end{equation} 
    for different initial phase differences $\Delta\Phi$ between the orthogonal modes of
    line Fe I 5250.65{\AA}. The averaging has been done over the $\mu$ values at which 
    observations are available. 
    This line is chosen, since it gives the tightest limits on $k$. The thick horizontal 
    line represents the value obtained from observations. Obviously above $k^2 = (13.8\, 
    \mbox{km})^2$ the curve obtained from theory always lies below the observed value, hence 
    ruling out such $k$ values. 
    
    For comparison with literature values it is useful to set limits on the 
    $\ell_{\odot}$-parameter in Moffat´s NGT, since earlier work has concentrated on 
    constraining this theory. Using the profile difference technique we obtain $\ell_{\odot}^2 
    < (178\, \mbox{km})^2$ measured in the line Fe I 5250.65{\AA}, compared with the 
    previous tightest upper limit of $(305\,\mbox{km})^2$.
  
  \subsection{Stokes asymmetry technique}
\begin{figure}[h]     
      \hspace*{1cm}\resizebox{3.5in}{1.745in}{\includegraphics{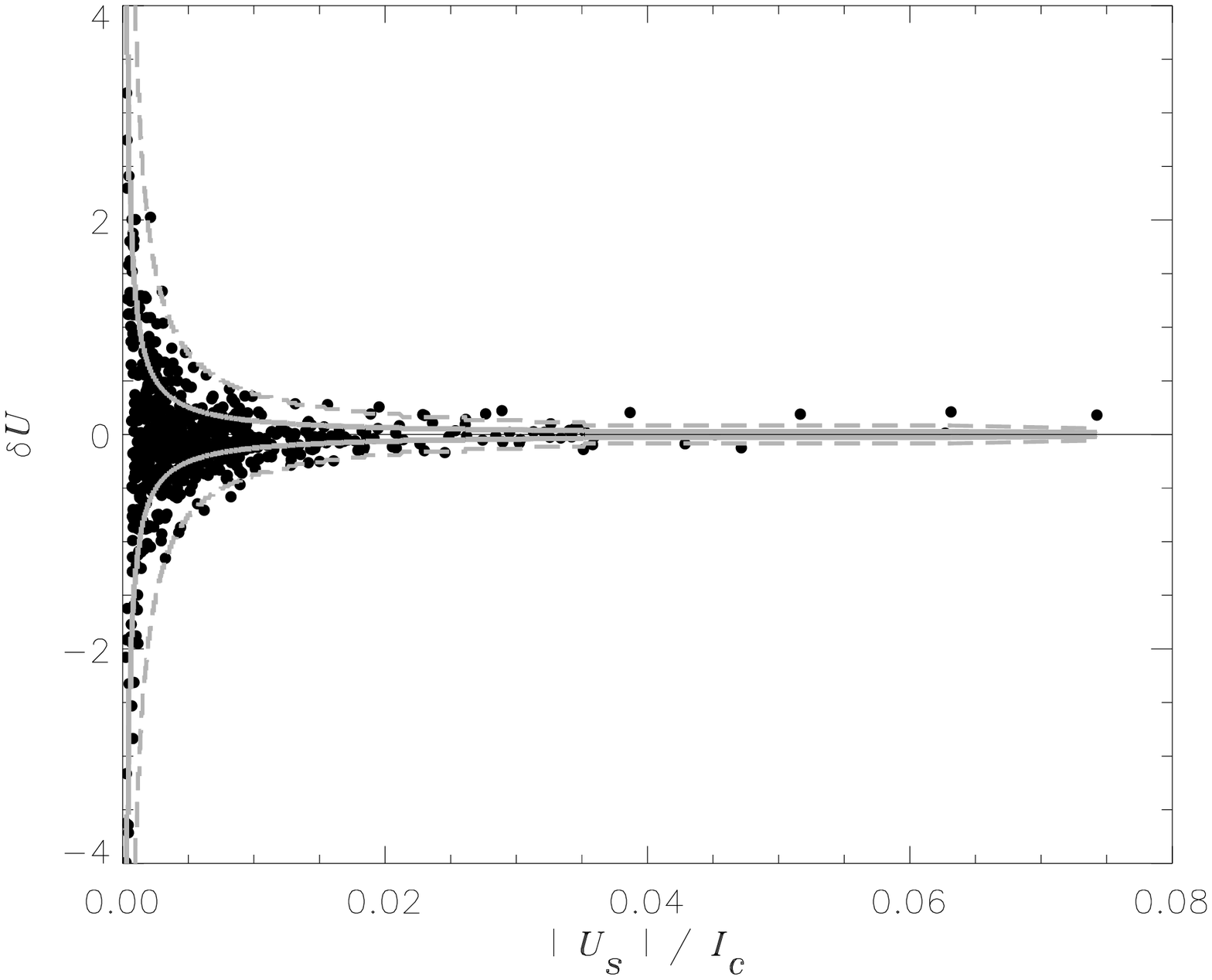}}
      \vspace{0.3cm}
      \hspace*{1cm}\resizebox{3.5in}{1.745in}{\includegraphics{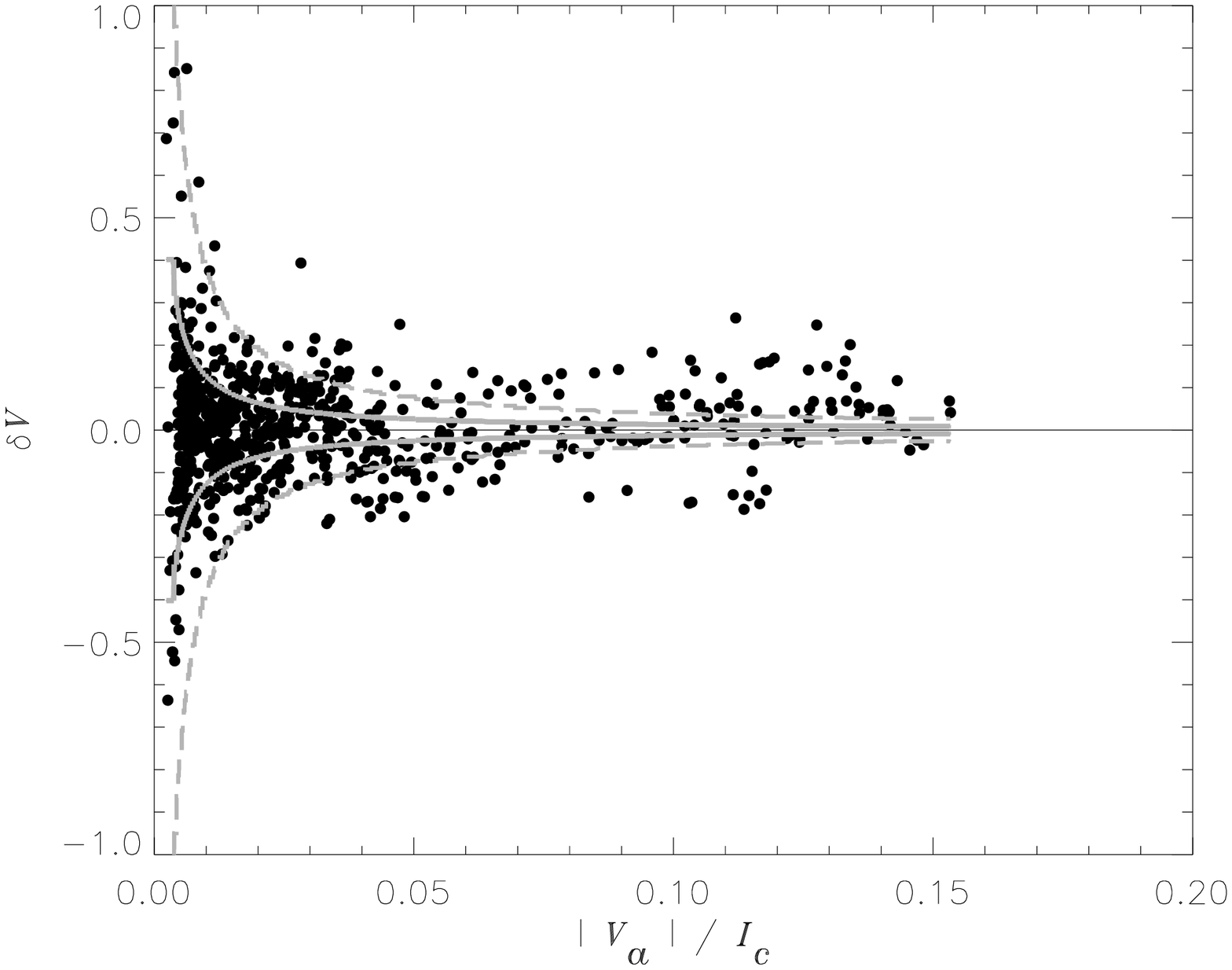}}
%      \vspace{0.5cm}
      \caption{Measured amplitude asymmetries for Stokes $U$ and Stokes $V$. }
      \label{sat1}
    \end{figure}

\begin{figure}[t]     
      \hspace*{1cm}\resizebox{3.8in}{3.6in}{\includegraphics{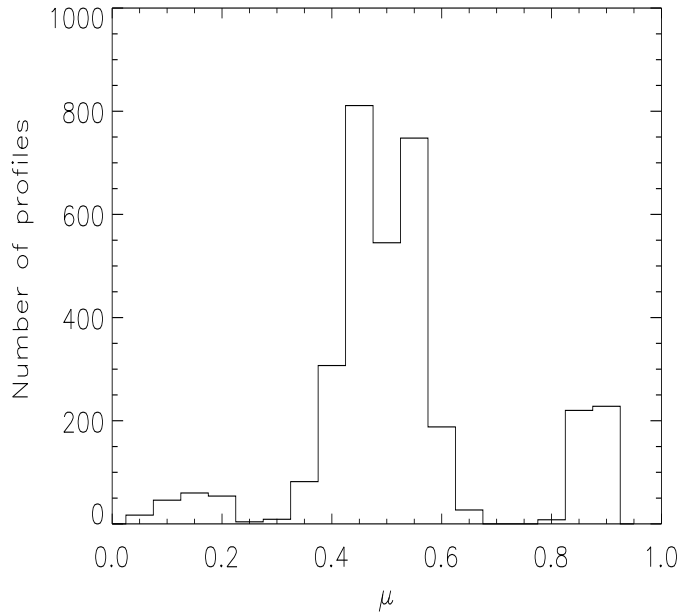}}      
      \caption{Histogram of the number of recorded profiles as a function of $\mu$.}    
      \label{histo}
    \end{figure}

    A measure of the asymmetry of a Stokes profile is given by the ratio 
    $\delta V = V_{\fs}/V_{\fa}$, respectively $\delta U = U_{\fa}/U_{\fs}$ 
    \cite{sos84,mp97})
    In Fig.~\ref{sat1} we plot these quantities vs. $|V_{\fa}|$ and $|U_{\fs}|$, respectively. 
    Each point in these plots refers to a Stokes profile of the Fe I 5250.65{\AA} line. 
    Although our observations cover a range of $\mu$ values, Fig.~\ref{sat1} is 
    similar to corresponding figures based on data obtained near $\mu = 1$ 
    \cite{gd96,mp97}, where the influence of gravitational birefringence vanishes for 
    symmetry reasons. For large amplitudes ($V_{\fa}$, $U_{\fs}$) the relative asymmetry 
    ($V_{\fs}/V_{\fa}$, $U_{\fa}/U_{\fs}$) is small, while for weaker profiles it shows 
    an increasingly large spread. For the weaker profiles this spread is mainly due to 
    noise as can be judged from the solid and dashed curves in Fig.~\ref{sat1}, which 
    outline the $1\sigma$ and $3\sigma$ spread expected due to photon noise, respectively. 
    The curves reveal that Stokes profiles with amplitudes ($V_{\fa}$, $U_{\fs}$) 
    smaller than one percent of the continuum intensity are so strongly affected by noise that 
    they are of little use for the present purpose. This leaves us with 1966 individual profiles 
    for further analysis. In Fig.~\ref{histo} we plot a histogram of the number of these profiles 
    as a function of $\mu$. The distribution is uneven, beeing determined by the position on the 
    solar disc of magnetic features at the times of the observations.

    The further analysis is made more complicated by the fact that a $V_{\fs}$ and a $U_{\fa}$
    signal can be produced not just by gravitational birefringence, but also by radiative
    transfer processes acting in the dynamic solar atmosphere, as described in Sect. 3.
    To circumvent this problem we consider all profiles satisfying the criterion that
    $|V_{\fa}|$ or $|U_{\fs}| \geq 0.01$. For all these profiles $|\delta V_{\obs}| < 0.7$
    and $|\delta U_{\obs}| < 0.6$. A similar picture is also obtained at $\mu = 1$. 
    Hence one way to limit $k^2$ is to require $|\delta V_{\src}| < 1$ and
    $|\delta U_{\src}| < 1$ for all $\mu$. This condition is strengthened
    by the fact that $|\delta V_{\obs}|$ and $|\delta U_{\obs}|$ decrease with decreasing $\mu$,
    \cite{sta87,mp97}, whereas gravitational birefringence increases towards the limb, 
    so that one would expect exactly the opposite behaviour if gravitational birefringence 
    had a significant effect on the Stokes $V$ or $U$ profiles.
  
\begin{figure}[t]     
      \centerline{\resizebox{.2in}{1.65in}{\includegraphics[angle=90]{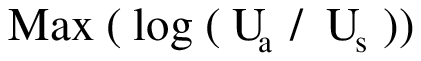}}\hspace{-0.4cm}
      \resizebox{3.5in}{2.8in}{\includegraphics[angle=90]{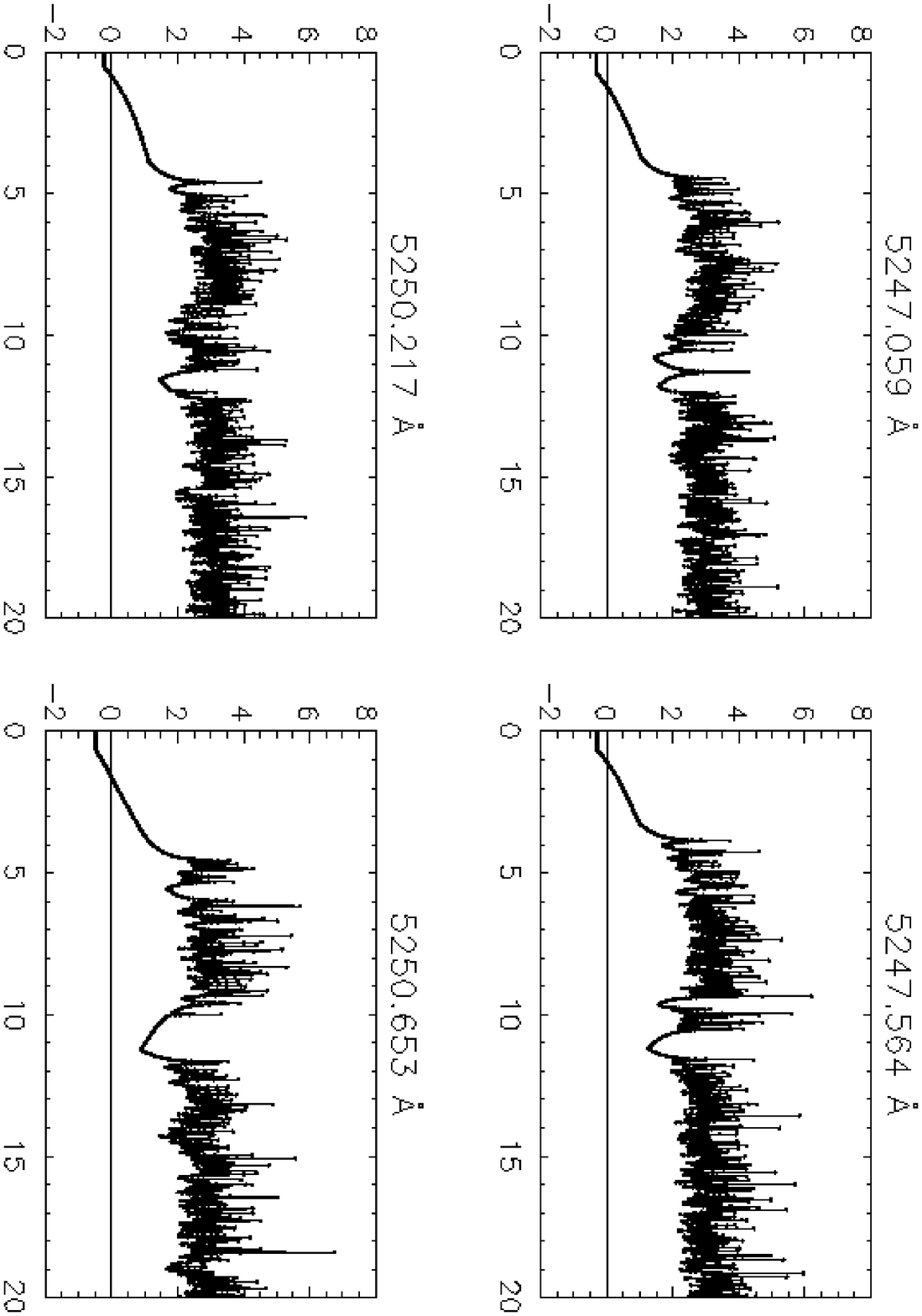}}}
      \vspace{0.3cm}
      \caption{Maximum of $(U_{\asc}/U_{\ssc})$, on a logarithmic scale, values for all 4 spectral 
               lines vs. $k$.}
      \label{max}
    \end{figure}

    In Fig.~\ref{max} we plot the maximum $(U_{\asc}/U_{\ssc})$ value predicted for 
    each of the 4 spectral lines, based on all analysed data, vs. $k^2$. 
    The horizontal line represents $\log(U_{\asc}/U_{\ssc} = 1)$, a limit 
    above which this ratio is not observed at $\mu \approx 1$. Clearly, as $k^2$ 
    increases $(U_{\asc}/U_{\ssc})_{\fm}$ initially remains almost 
    equal to $(U_{\aob}/U_{\sob})_{\fm}$, but begins to increase 
    for $k^2 \gsim (1\,\mbox{km})^2$, becoming $\gsim 10$ at $k^2 < (2\, 
    \mbox{km})^2$ and finally oscillating around $(U_{\asc}/U_{\ssc})
    _{\fm}$ of 100 - 1000. All 4 spectral lines exhibit a similar behaviour, implying 
    that the influence of noise is very small. The largest effect of gravitational birefringence 
    is exhibited by the two most strongly Zeeman split lines, Cr I 5247.56{\AA}, Fe I 5250.22 
    {\AA}, which also produce the largest Stokes $V$ and $U$ signals, while the lines with 
    the smallest splitting, Fe I 5250.65 {\AA} provides the weakest limit.  
    
    It is in principle sufficient to limit $k$ by requiring that none of the observed
    spectral lines has $(U_{\asc}/U_{\ssc}) > 1$ within the range of allowed 
    $k$ values. This gives $k^2 < (0.9\, \mbox{km})^2$. A limit obtained 
    similarly from Stokes $V$ is both larger $k^2 < (3.42\, \mbox{km})^2$ and less 
    reliable, since we cannot completely rule out $V_{\ssc}/V_{\asc} > 1$ to be present, 
    although we expect such profiles to be very rare, among large amplitude profiles.
  
    For Moffat's NGT this technique gives $\ell^2_{\odot} < (57,1\, \mbox{km})^2$,
    implying a 28-fold reduction in $\ell^2_{\odot}$ compared to previous work.
    
\begin{figure}[t]   
  \vspace{-0.3cm}  
    \hspace*{1cm}\resizebox{3.8in}{2.5in}{\includegraphics[angle=90]{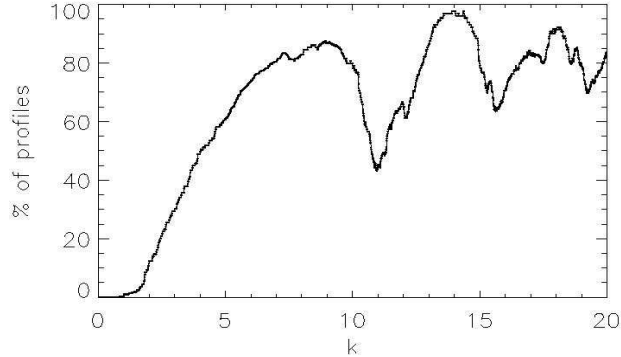}}
    \caption{Fraction of profiles with $(U_{\asc}/U_{\ssc}) > 1$ for NGT (top) and MAG (bottom).}
    \label{umag}
  \end{figure}

    An alternative test is to determine the fraction of profiles with $(U_{\asc}/
    U_{\ssc}) > 1$ (Fig.~\ref{umag}). Fig.~\ref{umag} reveals that initially no $U$ profile satisfies
    the criterion, above $k^2_{\odot} = (1.33\, \mbox{km})^2$ 1.7 \% of the profiles does. 
    This number keeps increasing with $k$, before finally oscillating around 70\% 
    at large $k$. Thus 10\% of all data points have $(U_{\asc}/U_{\ssc}) > 1$ for 
    $k^2 = (1.955\, \mbox{km})^2$, 20\% for $k^2 = (2.43\, \mbox{km})^2$.
    We are not aware of any solar observations of Stokes $U$ with $\delta U > 1$ for which
    instrumental cross-talk is negligible \cite{sal92,sku90}. $k^2 < (2.5\, \mbox{km})^2$ 
    is thus a very conservative upper limit, deduced from this criterion.

    For $\ell^2$, we found that 10\% of all data points have $(U_{\asc}/U_{\ssc}) > 1$ for 
    $\ell^2_{\odot} = (74.0\, \mbox{km})^2$ and 20\% for $\ell^2_{\odot} = (79.4\, \mbox{km})^2$.

\section{Discussion and Conclusions}
  New couplings between electromagnetism and gravity introduced within the framework of
  Metric-affine Gravity (MAG) theories lead to observable effects, in particular to
  space-time becoming birefringent in the presence of a gravitational field. By constraining 
  the level of this birefringence we can constrain the strength of the coupling between electromagnetism
  and gravity within this particular framework. A birefringence of space-time is also expected
  in a more general context, so that such a constraint is also of wider significance.
  
  Using two techniques (the Stokes asymmetry technique and the new profile difference technique
  applied to solar data) we have imposed stringent constraints on the coupling constant $k$
  introduced by MAG. In order to judge how stringent the constraints imposed by these techniques
  are, we have also redone the analysis for Moffat's NGT. The new techniques improved previous 
  limits on $\ell^2$ given by Solanki \& Haugan (1996) \cite{sh96} by nearly one order of 
  magnitude. It will be difficult to set much tighter limits on gravitational birefringence
  than those found here using solar data in the visible spectral range. To obtain a significant 
  improvement one would need to observe at shorter wavelengths. The spectral line at the shortest
  wavelength that is strong enough to provide a hope of detecting Stokes $U$ and $V$ at
  sufficient $S/N$ is Ly$\alpha$ at 1216{\AA}. The maximum gain that one could expect relative 
  to the current analysis is a factor of
  \begin{equation}
   \frac{\lambda_{\mbox{\footnotesize{visible}}}}{\lambda_{\mbox{\footnotesize{Ly}}\alpha}} = \frac{5250}{1216} = 4.32
  \end{equation}
  Another possibility is to consider astronomical objects with a stronger gravitational field
  and well defined sources of polarization. In a parallel paper \cite{p03} we employ magnetic
  white dwarfs for this purpose.


\begin{thebibliography}{99}
  \bibitem{dicke} R.H. Dicke, Experimental relativity. In ''{\em Relativity, Groups
                  and Topology}'', ed. C. DeWitt and B. DeWitt, pp.165-313, Gordon and
		  Breach, New York (1964)
  \bibitem{ni77}  W.-T. Ni, Phys. Rev. Lett. {\bf 38}, 301 (1977).  
  \bibitem{ni84}  W.-T. Ni, in {\it Precision Measurements and Fundamental Constants II}, 
                  edited by B.N. Taylor and W.D. Phillips, U.s. National Bureau of Standards 
                  Publication 617 (U.S. GPO, Washington D.C., 1984).  
  \bibitem{sh96}  S.K. Solanki, M.P. Haugan, Phys. Rev. D {\bf 53}, 997 (1996) 
  \bibitem{s99}   S.K. Solanki, M.P. Haugan, R.B. Mann, Phys. Rev. D {\bf 59}, 047101 (1999).
  \bibitem{mof79} J.W. Moffat, Phys. Rev. D {\bf 19}, 3554 (1979); {\bf 19},
                  3562 (1979); {\bf 35}, 3733 (1987); {\em Gravitation: A Banff
		  Summer Institute}, edited by R.B. Mann and P. Wesson (World
		  Scientific, Singapore, 1991).
  \bibitem{gab91} M.D. Gabriel, M.P. Haugan, R.B. Mann, J.H. Palmer, Phys. 
                  Rev. D. {\bf 43}, 308 (1991). 		  
  \bibitem{gab91a} M.D. Gabriel, M.P. Haugan, R.B. Mann, J.H. Palmer, Phys.
                  Rev. D. {\bf 43}, 2465 (1991).
  \bibitem{gab91b} M.D. Gabriel, M.P. Haugan, R.B. Mann, J.H. Palmer, Phys. Rev. Lett. 
                  {\bf 67}, 2123 (1991).
  \bibitem{hk95}  M.P.Haugan and T.F. Kauffmann, Phys. Rev. D {\bf 52}, 3168 (1995).
  \bibitem{d92}   T. Damour, S. Deser, J. McCarthy, Phys. Rev. D {\bf 45}, R3289 (1992).
  \bibitem{c98}   M. Clayton, L. Demopoulos, J. Legare, GRG {\bf 30}, 1501 (1998).
  \bibitem{h95}   F.W. Hehl, J.D. McCrea, E.W. Mielke, Y. Ne'eman, Phys. Repts. 258, 1 (1995).
  \bibitem{r03}   G.F. Rubilar, Y.N. Obukhov, F.W. Hehl, Class. Quantum Grav. {\bf 20} (2003)
                  L185 ({\em Preprint} {\tt gr-qc/0305049}).
  \bibitem{ih03}  Y. Itin, F.W. Hehl, {\em Preprint} {\tt gr-qc/0307063}		  
  \bibitem{p97}   R.A. Puntigam, C. Laemmerzahl, F.W. Hehl, Class. Quantum Grav. {\bf 14}, 
                  1347 (1997). 
  \bibitem{t95}	  R. Tresguerres, Z. Phys. C {\bf 65}, 347 (1995).
  \bibitem{t95a}  R. Tresguerres, Phys. Lett. A  {\bf 200}, 405 (1995).  
  \bibitem{edm}   A.R. Edmonds; {\em ''Angular Momentum in Quantum Mechanics''}; 
                  Princeton University Press, Princeton 1974  
  \bibitem{sos84} S.K. Solanki, J.O. Stenflo, Astron. Astrophys. {\bf 140} 185 (1984)
  \bibitem{gd89}  U. Grossmann-Doerth, M. Sch\"ussler, S.K. Solanki, Astron. Astrophys. {\bf
                  221} 338 (1989) 
  \bibitem{sgs99} U. Grossmann-Doerth, M. Sch\"ussler, M. Sigwarth, O. Steiner, Astron. 
                  Astrophys. {\bf 357} 351 (2000)  		  
  \bibitem{gd96}  U. Grossmann-Doerth, C.U. Keller, M. Sch\"ussler, Astron. Astrophys. {\bf 315}
                  610 (1996)
  \bibitem{mp97}  V. Mart\'{\i}nez Pillet, B.W. Lites, A. Skumanich, Astrophys. J. {\bf 474},
                  810 (1997)
  \bibitem{sea93} S.K. Solanki, I. Ruedi, D. Rabin, ASP {\bf 46}, 534 (1993)		  
  \bibitem{sal92} J. S\'{a}nchez Almeida, B.W. Lites, Astrophys. J. {\bf 398}, 359 (1992)
  \bibitem{plo01} S.R.O. Ploner, M. Sch\"ussler, S.K. Solanki, A.S. Gadun, 2001, M.
                  Sigwarth(ed.), {\em Advanced Solar Polarimetry - Theory, Observation and 
		  Instrumentation. 20TH NSO/Sac Summer Workshop, ASP Conference Proceedings, 
		  Vol. 236. Edited by Michael Sigwarth.} ISBN: 1-58381-073-0, 2001, p.363
  \bibitem{pov95} H.P. Povel, Opt. Eng., {\bf 34}, 1870 (1995)
  \bibitem{sea86} S.K. Solanki, J.O. Stenflo, Astron. Astrophys. {\bf 170}, 120 (1986)
  \bibitem{smw91} J. S\'{a}nchez Almeida, V. Mart\'{\i}nez Pillet, A.D. Wittmann, Solar Phys. {\bf 134},
                  1 (1991)
  \bibitem{smk95} J. S\'{a}nchez Almeida, V. Mart\'{\i}nez Pillet, F. Kneer, Astron. Astrophys. 
                  {\bf 113}, 359 (1995)		    
  \bibitem{ber97} P.N. Bernasconi, Ph.D. Thesis, ETH, Zuerich (1997)
  \bibitem{gap97} A.M. Gandorfer, H.P. Povel, Astron. Astrophys. {\bf 328}, 381 (1997)
  \bibitem{pov98} H.P. Povel, in: Magnetic Fields Across the Hertzsprung-Russell Diagram, ASP 
                  Conference Proceedings Vol. 248. Edited by G. Mathys, S. K. Solanki, and 
		  D. T. Wickramasinghe. ISBN: 1-58381-088-9. San Francisco: Astronomical Society 
		  of the Pacific, 2001.
  \bibitem{fas98} M. Fligge, S.K. Solanki, ASP Conf. Ser. 154, {\em The Tenth Cambridge Workshop 
                  on Cool Stars, Stellar Systems and the Sun}, p.833	 
  \bibitem{gd91}  U. Grossmann-Doerth, M. Sch\"ussler, S.K. Solanki, Astron. Astrophys.
                  {\bf 249}, 239 (1991)
  \bibitem{sta87} J.O. Stenflo, S.K. Solanki, J.W. Harvey, Astron. Astrophys. {\bf 171}, 305 (1987)
  \bibitem{sku90} A. Skumanich, B. Lites, {\em Proceedings of the Eleventh National Solar Observatory
                  /Sacramento Peak Summer Workshop, Sunspot, New Mexico, 311 (1990)}   
  \bibitem{p03}   O. Preuss, M.P. Haugan, S.K. Solanki, S. Jordan, {\em Astronomical search for 
                  evidence of new physics: Limits on gravity-induced birefringence from the 
		  magnetic white dwarf RE J0317-853}, accepted for publication in Phys. Rev. D (2003)
\end{thebibliography}
\end{document}